\documentclass[lettersize,journal]{IEEEtran}
\usepackage{amsmath,amssymb,amsfonts}
\usepackage{algorithmic}
\usepackage{array}
\usepackage[caption=false,font=normalsize,labelfont=sf,textfont=sf]{subfig}
\usepackage{textcomp}
\usepackage{stfloats}
\usepackage{url}
\usepackage{verbatim}
\usepackage{graphicx}
\usepackage{subeqnarray}
\usepackage{cases}
\usepackage[T1]{fontenc}
\usepackage[usenames, dvipsnames]{color}
\def\BibTeX{{\rm B\kern-.05em{\sc i\kern-.025em b}\kern-.08em
    T\kern-.1667em\lower.7ex\hbox{E}\kern-.125emX}}
\usepackage{balance}
\begin{document}
\title{A Comparative Study on the Convergence Rate of Two Online Quantum State Reconstruction Algorithms}

\author{~\IEEEmembership{Shuang Cong, Senior Member, IEEE, and  Weiyi Qin}

\thanks{This paper was produced by the Department of Automation, University of Science and Science of China, Hefei, 230027.  This work was supported in part by the National Natural Science Foundation of China under Grant No. 62473354.}
\thanks{Manuscript received April 29, 2024; revised August 16, 2021.}}

\markboth{Journal of \LaTeX\ Class Files,~Vol.~18, No.~9, September~2020}%
{How to Use the IEEEtran \LaTeX \ Templates}

\maketitle

\begin{abstract}
   In this paper, the convergence rates of two algorithms for the online quantum states reconstruction with Gaussian measurement noise in continuous weak measurement are studied, one is the online proximal gradient-based alternating direction method of multipliers (OPG-ADMM) algorithm, and another is Kalman fitering-based quantum state estimation (KF-QSE) algorithm. For the OPG-ADMM algorithm, by defining the loss function of the optimization function and the constraint condition in the times $T$ tracking process, the convergence rate theorem of the two loss functions is obtained and proved. Then, the convergence order of the normalized distance of the density matrix under the OPG-ADMM algorithm is derived from the conclusion of the theorem. For the KF-QSE algorithm, after defining the loss function of the optimization function, the theorem of the convergence order of the loss function is investigated. Then, the convergence order of the normalized distance of the KF-QSE algorithm is deduced from the conclusion of the theorem. Finally, in the numerical simulation experiments, we use the normalized distance of density matrix as the indicator and use two algorithms for online reconstruction of the 4-bit quantum system. The derived performance of algorithm convergence rates are verified by the comparison and analysis of the results.
\end{abstract}


\section{Introduction}
\IEEEPARstart{Q}{uantum} state reconstruction, also known as quantum state tomography or quantum state estimation, is an important prerequisite for realizing high-precision state feedback control of quantum systems and quantum state preperation in quantum experiments \cite{sayrin2011real}. Online algorithm refers to a class of algorithms that use data serialization to process real-time data, which is needed to use in the real quantum system experiments. The goal of online quantum state reconstruction is to reconstruct quantum states that change dynamically over time \cite{gyongyosi2017quantum}. In 2005, Silberfar proposed the continuous weak measurement (CWM) theory \cite{silberfarb2005quantum}, which provided a theoretical basis for online quantum state reconstruction. They first developed an online maximum likelihood (OML) reconstruction using the ML method \cite{silberfarb2005quantum,smith2006efficient}. Ralph et al. derived a complete Bayesian estimation algorithm for frequency tracking and parameter estimation of single-qubit system \cite{ralph2011frequency}. 
Shanmugam and Kalyani studied unrolling SVT to obtain computationally efficient SVT for N-qubit quantum state tomography \cite{Shanmugam2023tomography}. Li et al. researched the Low-rank positive semidefinite matrix recovery from corrupted rank-one measurements \cite{Li2017Low-Rank}.

The research on online optimization algorithms and their convergences have provided theoretical guidance and technical tools for the experiment realization \cite{hazan2007logarithmic,zinkevich2003online}. Hosseini et al. designed an online optimization algorithm with two optimization variables based on the ADMM algorithm and proved the algorithm convergence rate theorems in two cases: non-strongly convex and strongly convex optimization functions \cite{hosseini2014online}. By dividing the optimization function according to differential properties, Dixit et al. designed the non-differentiable online gradient descent algorithm (OGD) and proved the convergence rate theorem \cite{huang2013algorithm}. Zhang et al. designed an online ADMM algorithm based on the original proximal gradient descent for online convex optimization problems with time-varying constraints.\cite{zhang2021online}. Wang and Banerjee studied the composite objective optimization problem when the optimization function with and without Bregman divergence. They proposed two online ADMM algorithms and proved the corresponding convergence rate theore \cite{wang2012online}.

 In order to solve the problem of time consuming in online quantum state reconstruction, online optimization algorithms must be used in the online quantum state reconstruction. In 2019,
Zhang et al. developed the convex optimization algorithm for quantum state reconstruction based on the ADMM algorithm for different types of interference in quantum systems \cite{zhang2019efficient}. In 2020, Tang et al. proposed a real-time density matrix reconstruction algorithm for single-qubit stochastic quantum systems based on CWM \cite{ya2020line}.
 In 2021,  Cong et al. proposd a new information acquisition and processing method for online quantum state reconstruction based on continuous weak measurements and compressed sensing theory \cite{cong2021line}. In the same year, Zhang et al. proposed an online proximal gradient-based alternating direction method of multipliers (OPG-ADMM) algorithm  \cite{zhang2021eff}, They also proposed a Kalman fiteingr-based quantum state estmation (KF-QSE) algorithm  \cite{cong2021}.

\par Up to now, there have been little studies on  convergence and convergence rate of online quantum state reconstruction algorithms. We have studied the convergence of online optimization algorithm of the quantum state filter with disturbance and noise \cite{cong2023}. This paper studies the the convergence rate of OPG-ADMM algorithm and the KF-QSE algorithm for online quantum state reconstruction. For the OPG-ADMM algorithm, by defining the loss function of optimization function and constraints condition in the tracking process, we propose the theorem of convergence order of two loss function; In the proof process, we first give the lemma to prove that the optimization function in each sampling time is bounded. Apply the results in the lemma to prove that the upper bound of the loss function of the optimization function continues to decrease as the tracking process proceeds, and the theorem gives the convergence order of this loss function. Finally, the convergence order of the quantum state density matrix normalized distance under the OPG-ADMM algorithm is derived from the conclusion of the loss function convergence order theorem. For the KF-QSE algorithm, first define the loss function of the optimization function, and give the theorem of the convergence order of the loss function; For the proof process, we prove the lemma that the distance of optimization function between the reconstructed value and optimal value continues to decrease as the tracking process proceeds. Based on this conclusion, we use another lemma to prove that the sum distance of optimization function between the reconstructed value and optimal value during the subsequent process is bounded. Based on the conclusions of lemmas, we prove the convergence order of the loss function of the optimization function. Convergence order of the quantum state density matrix normalized distance under the KF-QSE algorithm is derived from the conclusion of the loss function convergence order theorem; Finally, in the simulation experiment, the density matrix normalized error is taken as the indicator, the algorithm is applied to the online reconstruction of 4-bit quantum system, and the results of the theoretical analysis are verified.

\section{Preparation work}
\subsection{Online quantum state reconstruction problem}
An open $n$-qubit system can be described by a stochastic master equation in Schrödinger picture \cite{yamamoto2017entanglement}. Based on the continuous weak measurement theory, stochastic master equation can be transformed into the corresponding approximate discrete evolution equation \cite{Harraz2019}:
\begin{equation}\label{E1}
  {\rho _k} = \sum\limits_{i = 1}^{{2^n}} {{A_i}(\Delta t){\rho _{k - 1}}{A_i}{{(\Delta t)}^\dag }} {\rm{    }},\ \ k = 2,3, \ldots ,N
\end{equation}
where $k = 2,3, \ldots ,N$ denotes the sampling times, ${A_i}(\Delta t) \in {C^{d \times d}}(i = 1, \ldots ,2n)$ is evolution operators of quantum system. $\rho  \in {C^{d \times d}}$ represents the Density matrix variables that need to be reconstructed and satisfies the quantum state constraint $C: = \{ \rho  \succ 0,tr(\rho ) = 1,{\rho ^\dag } = \rho \} $.
\par By using the sliding window method to constructing the measurement sequence $b_k$, \eqref{E1} is transformed into a double objective convex optimization problem with the Gaussian noise in the measurement \cite{zhang2021eff} :
\begin{equation}\label{E2}
  \begin{array}{l}
    \mathop {\min }\limits_{\rho ,e} \frac{1}{{2\gamma }}\left\| e \right\|_2^2 + \left\| {\rm{vec}(\rho  - {\rho _{k - 1}})} \right\|_2^2\\
    s.t.\ \ {\rm{    }}\par
    {{{\cal A}}_k}\rm{vec}(\rho ) + e = {b_k}{\rm{ }}\\
    \rho \succ 0,tr(\rho ) = 1,{\rho ^\dag } = \rho 
    \end{array}
\end{equation}
where $\gamma > 0$ denotes a regular parameter. ${e_k} \in {R^k}(l < k)$ is Gaussian measurement noise of measurement sequence whose mean is 0. $b_k$ is a data set of measurement sequence with length $l>0$ based on the sliding window method \cite{zhang2020online}, defined as:
\[
  {b_k} = \left\{ \begin{array}{l}
    {({y_1}, \ldots ,{y_{k - 1}},{y_k})^T},{\rm{         }}k < l\\
    {({y_{k - l + 1}}, \ldots ,{y_{k - 1}},{y_k})^T},{\rm{    }}k \ge l
    \end{array} \right.
\]
where $y_i$ is the measurement value of the quantum system at each sampling time in Schrödinger picture. $y_i$ is obtained by using macroscopic instruments, so it is generally equal to the average value of the measurement operator acting on the quantum state, which defined as:
\[{y_i} = tr(M_1^\dag {\rho _i}) = \rm{vec}{({M_1})^\dag }\rm{vec}({\rho _i})\]
where $M_k$ is the measurement operator, defined as:
\[{M_k} = \sum\limits_{i = 1}^{{2^n}} {{M_i}{{(\Delta t)}^\dag }{M_{k - 1}}{M_i}(\Delta t)}\]
where $M_i(\Delta t)$ denotes the weak measurement operators of an $n$-qubit system \cite{zhang2020online}.\\
$\mathcal{A}_k$ is the sampling matrix corresponding to $b_k$, which is defined as:
\[
  {{\cal A}_k} = \left\{ \begin{array}{l}
    {(\rm{vec}({M_k}), \ldots ,\rm{vec}({M_2}),\rm{vec}({M_1}))^\dag },{\rm{         }}k < l\\
    {(\rm{vec}({M_l}), \ldots ,\rm{vec}({M_2}),\rm{vec}({M_1}))^\dag },{\rm{          }}k \ge l
    \end{array} \right.
\]
\par For the convex optimization problem \eqref{E2}, the variable $\rho_k$ corresponding to the optimization function is $\|\rm{vec}(\rho  - {\rho _{k - 1}}){_2}\|_2$; For the  variable $e_k$, its corresponding optimization function is $\frac{1}{{2\gamma }}\|e\|_2^2$: The constraint conditions are ${{{\cal A}}_k}\rm{vec}(\rho ) + e = {b_k}$.
\subsection{Online optimization algorithm OPG-ADMM}
By using OPG-ADMM algorithm \cite{zhang2021eff}, problem \eqref{E2} can be decomposed into two subproblems by setting dual variables and cross-updating optimization variables:
\begin{subnumcases} {\label{E3}}
  {\rho _k} = \arg\mathop {\min }\limits_\rho  \frac{\alpha }{2}\left\| {{{{\cal A}}_k}\rm{vec}(\rho ) + {e_{k - 1}} - {b_k} - \frac{{{\lambda _{k - 1}}}}{\alpha }} \right\|_2^2 \\ \label{E3a}
  \ \ \ \ \ \ + \frac{1}{{2\eta }}\left\| {\rm{vec}(\rho  - {\rho _{k - 1}})} \right\|_{{P_k}}^2{\rm{      }}\notag\\ 
  {e_k} = \arg\mathop {\min }\limits_e \{ \frac{1}{{2\gamma }}\left\| e \right\|_2^2 \\
    \ \ \ \ \ \ + \frac{\alpha }{2}\left\| {{{{\cal A}}_k}\rm{vec}({{\rho }_k}) + e - {b_k} - \frac{{{\lambda _{k - 1}}}}{\alpha }} \right\|_2^2\} {\rm{                           }}\notag\\ \label{E3b}
    {\lambda _k} = {\lambda _{k - 1}} - \alpha ({{{\cal A}}_k}\rm{vec}({\rho _k}) + {e_k} - {b_k}){\rm{                                                           }}\label{E3c}
\end{subnumcases}

for subproblem (3), when the tracking process is $T$ times, the parameter selection condition that makes the algorithm converge is
\begin{equation}\label{E4}
  \begin{array}{l}
    a.{\rm{  }}{P_k} = \tau I - \alpha \eta {\cal A}_k^{\dagger}{{\cal A}_k} \succ 0\\
    b.{\rm{  }}\alpha {\rm{ = }}\sqrt T ,\eta  = \frac{\tau }{{\alpha {\sigma _m}}},{\sigma _m} = {\lambda _{\max }}({\cal A}_{}^{\dagger}{{\cal A}_{}})\\
    \gamma  > 0,\tau  > 0
    \end{array}
\end{equation}
for subproblem (3a), it could update with first-order gradient by ignoring quantum state constraints first:
\begin{equation}\label{E5}
  \begin{gathered}
    \rm{vec}({\tilde \rho _{k - 1}}) = \rm{vec}({\rho _{k - 1}}) - \frac{\alpha }{{{\tau _1}}}{\cal A}_k^\dag ({{\cal A}_k}\rm{vec}({\rho _{k - 1}}) +\\
    {e_{k - 1}} - {b_k} - \frac{{{\lambda _{k - 1}}}}{\alpha })
  \end{gathered}
\end{equation}
where $\tilde \rho_{k-1}$ is a temporary variable only used in \eqref{E5}. $\rho_k$ can be obtained by projection ${\tilde \rho _{k - 1}}$ into the solution space of quantum state constraints through ISTA algorithm \cite{daubechies2004iterative}.
\par For subproblem (3b), it could be updated directly by the first-order gradient.
\par If the reconstructed variables $\{\rho_k, e_k\}$ are updated according to the subproblems in \eqref{E3} under the parameter selection conditions in \eqref{E4}, the value of reconstructed variables $\{\rho_k, e_k\}$ can converge to the optimal value $\{\rho_k^*,e_k^*\}$.
\subsection{Online optimization algorithm KF-QSE}
The KF-QSE online algorithm updates the reconstructed variable $\rho_k$ by constructing a Kalman filter matrix instead of first-order stochastic gradient and brings constraints ${{{\cal A}}_k}\rm{vec}({\rho _k}) + {e_k} = {b_k}$ into the optimization function $\frac{1}{{2\gamma }}\|e\|_2^2$ to solve the convex optimization problem \eqref{E2}. The KF-QSE algorithm can be divided into the following two steps at sampling times $k$:\\
\textbf{Step 1:} Define ${K_k} = A_k^\dag {(I + {A_k}A_k^\dag )^{ - 1}}$ as Kalman filter matrix. then, Gradient descent using $K_t$:
\begin{equation}\label{E6}
  \rm{vec}({\hat \rho _{k}}) = \rm{vec}({\rho _{k - 1}}) - {K_k}({A_k}\rm{vec}({\rho _{k - 1}}) - {b_k})
\end{equation}
where ${\hat \rho _{k}}$ is a temporary value only used in \eqref{E6}.\\
\textbf{Step 2:} Considering the quantum state constraint $C$,the reconstructed density matrix $\rho_k$ can be obtained by solving the following convex optimization problem:
\begin{equation}\label{E7}
  \begin{array}{l}
    {\rho _{k}} = \arg \mathop {\min }\limits_\rho  \left\| {\rho  - \frac{{{{\hat \rho }_{k}} + \hat \rho _{k}^\dag }}{2}} \right\|_F^2\\
    {\rm{  }}s.t.\ \ {\rm{   }}\rho  \succ 0,tr(\rho ) = 1
    \end{array}
\end{equation}
problem \eqref{E7} can be solved using ISTA algorithm \cite{daubechies2004iterative}.
\par For KF-QSE algorithm, update $\tilde{\rho}_k$ according to \eqref{E6}, and update $\rho_{k+1}$ with ISTA algorithm according to problem \eqref{E7}, then the value of reconstructed variable $\rho_k$ can converge to the optimal value $\rho_k^*$.
\section{Main results}
\par The aim of online quantum reconstruction is to make the value of reconstructed variable $\rho_k$ track to the optimal value $\rho_k^*$ in the tracking process. The main work of this paper is to compare the convergence rates of KF-QSE and OPG-ADMM algorithms by deducing and proving the convergence rate theorems of the two algorithms and the convergence rate performances of real-time tracking dynamic quantum states.
\par The convergence rate of the algorithm can be measured by the convergence order. In the next two subsections, we will give the convergence rate of two algorithms shown by convergence order of the loss function in the $T$-times tracking process.
\subsection{Convergence rate of OPG-ADMM algorithm}
For the convenience of subsequent proofs and theorems, we define $p = \rm{vec}(\rho ) \in {R^{{2^{d \times d}}}}$ and replace all the $\rm{vec}(\rho)$ as $p$. Now, we get the equivalent form of subproblem \eqref{E3}
\begin{subnumcases} {\label{E8}}
  {p_k} = \arg\mathop {\min }\limits_\rho  \frac{\alpha }{2}\left\| {{{{\cal A}}_k}p + {e_{k - 1}} - {b_k} - \frac{{{\lambda _{k - 1}}}}{\alpha }} \right\|_2^2 \\
 \ \ \ \ \ \ + \frac{1}{{2\eta }}\left\| {p - {p_{k - 1}}} \right\|_{{P_k}}^2\} {\rm{          }}\notag\\
  {e_k} = \arg\mathop {\min }\limits_e \{ \frac{1}{{2\gamma }}\left\| e \right\|_2^2 \\
\ \ \ \ \ \ + \frac{\alpha }{2}\left\| {{{{\cal A}}_k}{p_k} + e - {b_k} - \frac{{{\lambda _{k - 1}}}}{\alpha }} \right\|_2^2\} {\rm{                        }}\notag\\
  {\lambda _k} = {\lambda _{k - 1}} - \alpha ({{{\cal A}}_k}{p_k} + {e_k} - {b_k}){\rm{                                                       }}
\end{subnumcases}
\par Define the optimization variable mean loss function $R_1(T)$ and the constraint mean loss function $R_2(T)$
\begin{equation}\label{E9}
  \begin{gathered}
    {R_1}(T) = \frac{1}{T}(\sum\limits_{k = 1}^T {(} \|{p_k}\|{_2} + \frac{1}{{2\gamma }}\|{e_k}\|_2^2) - \\
    \sum\limits_{k = 1}^T {(\|p_k^*\|{_2} + \frac{1}{{2\gamma }}\|e_k^*\|_2^2))} \\
    {R_2}(T) = \frac{1}{T}(\sum\limits_{k = 1}^T {\|{{{\cal A}}_k}{p_k} + {e_k} - {b_k}}\|_2^2)
  \end{gathered}
\end{equation}
under parameter condition \eqref{E4}, the convergence order of OPG-OADM algorithm for online quantum state reconstruction is given by theorem 1:\\
\textbf{Theorem 1:} For the OPG-OADM algorithm iteration sequence $\{ {p_k},{e_k},{\lambda _k}\} $, in a $T$-times tracking process, there are\\
\begin{equation}\label{E10}
  \begin{gathered}
    {R_1}(T) \le O(\frac{1}{{\sqrt T }})\\
    {R_2}(T) \le O(\frac{1}{{\sqrt T }})
    \end{gathered}
\end{equation}
where, $O(\frac{1}{\sqrt{T}})$ represents the term of $O(\frac{1}{\sqrt{T}})$ order.\\
\par At the same time, we adopt normalized distance $D(\rho _k^*,{\rho _k})$ as the performance index of online quantum state reconstruction:
\begin{equation}\label{E11}
  \begin{gathered}
  D(\rho _k^*,{\rho _k}) = \left\| {\rho _k^* - {\rho _k}} \right\|_F^2/\left\| {\rho _k^*} \right\|_F^2 \\
  = \left\| {p_k^* - {p_k}} \right\|_2^2/\left\| {p_k^*} \right\|_2^2 = D(p_k^*,{p_k})
  \end{gathered}
\end{equation}
the normalized distance $D(\rho _k^*,{\rho _k})$ reflects the distant under the $l_2$ norm between the value of reconstructed variable $\rho_k$ and the optimal value $\rho_k^*$. Value of $D$ is between (0,1), and the closer the value is to 0, the higher the similarity between the value of reconstructed variable and the optimal value.\\
\par Under the parameter condition \eqref{E4}, the order of normalized distance convergence of the OPG-OADM algorithm is given by theorem 2:\\
\textbf{Theorem 2:} For the OPG-OADM algorithm iteration sequence $\{ {p_k},{e_k},{\lambda _k}\} $, in a $T$-times tracking process, there is\\
\begin{equation}\label{E12}
  D(p_T^*,{p_T}) \le O(\frac{1}{{\sqrt T }})
\end{equation}
\subsection{Convergence rate of KF-QSE algorithm}
The same as the OPG-OADM algorithm, we replace $ve{\rm{c(}}\rho {\rm{)}}$ as $p$. Thus, the process of KF-QSE algorithm can be transformed into an equivalent form:\\
\textbf{Step 1:} Using gradient descent without considering constraints, equation \eqref{E6} becomes
\begin{equation}\label{E13}
  {\hat p_k} = {p_{k - 1}} - {K_k}({\mathcal{A}_k}{p_{k - 1}} - {b_k})
\end{equation}
\textbf{Step 2:} The results of gradient descent are projected to the constrained set by semidefinite programming(SDP), equation \eqref{E7} becomes
\begin{equation}\label{E14}
  \begin{gathered}
    {\rm{                                 }}{{\rm{p}}_k} = \arg{\min _p }\|p - \frac{\hat{p}_k+\hat{p}_k^\dagger}{2}\|_2^2\\
    s.t.\;\;\arg \rm{vec}(p)\succ 0,tr(\arg \rm{vec}(p)) = 1,\\
    \arg \rm{vec}{(p)^\dag } = \arg \rm{vec}(\rho )
    \end{gathered}
\end{equation}
\par The convergence rate of the algorithm can be obtained by using the order of the loss function of optimization function in $T$-times tracing process. The loss function is defined as 
\begin{equation}\label{E15}
  REG(T): = \mathop \sum \limits_{k = 1}^T [{h_k}({p_k}) - {h_k}(p_k^*)]
\end{equation}
where $h_k(\cdot)$ is
\begin{equation}\label{E16}
  \begin{gathered}
    {h_k}(p) = \frac{1}{{2\gamma }}\|{\mathcal{A}_k}p - {b_k}\|_2^2 + \|p - {p_{k - 1}}\|_2^2\\
    \end{gathered}
\end{equation}
\par $REG(T)$ represents the optimal function average distance between the reconstructed value $\rho_k$ and the optimal value $\rho_k^*$ in $T$-times tracking process.
\par Based on the defination of \eqref{E15}, the convergence order of the KF-QSE algorithm framework is given by theorem 3.\\
\textbf{Theorem 3:} For KF-QSE algorithm, the loss function $REG(T)$ in a $T$-times tracking process is satisfied:
\begin{equation}\label{E17}
  REG(T) \le O(\frac{1}{T}) + Er{r_T}
\end{equation}
where $Err_{T}$ is an error term that does not converge with the sampling times $T$.\\
\par We give the convergence order of normalized distance $D(\rho_k,\rho_k^*)$ to measure the theoretical performance of tracking dynamic quantum states. the order of normalized distance convergence of the KF-QSE is given by theorem 4:\\
\textbf{Theorem 4:} For KF-QSE, the normalized distance $D(p_T^*,p_T)$ is satisfied in a $T$-times tracking process:
\begin{equation}\label{E18}
  D(p_T^*,{p_T}) \le O(\frac{1}{T}) + Er{r_T}
\end{equation}
\section{Proof of the theorems}
\subsection{Proof of the theorem 1}
The proof process is as follows: We first prove by lemma 1 that the distance of the optimization function value between the values of reconstructed variables $\{ {p_k},{e_k}\}$ and the optimal values $\{ p_k^*,e_k^*\} $ is bounded at the sampling time $k$. By applying the result of Lemma 1, the upper bounds of the two loss functions decrease continuously in $T$ times is proved in theorem1, and the order of upper bounds is $O(\frac{1}{\sqrt{T}})$. Finally, theorem 2 is proved by the conclusion of theorem 1.\\
\textbf{Lemma 1:} For OPG-ADMM algorithm, $\forall k >0$, there is
\begin{equation}\label{E19}
  \begin{gathered}
    \|{p_k}\|{_2} + \frac{1}{{2\gamma }}\|{e_k}\|_2^2 - (\|p_k^*\|{_2} + \frac{1}{{2\gamma }}\|e_k^*\|_2^2)\\
     \le \frac{1}{{2\alpha }}(\|{\lambda_{k - 1}}\|_2^2 - \|{\lambda_k}\|_2^2) - \frac{\alpha }{2}{\|{{\cal A}}_k}{p_k} + {e_{k - 1}} - {b_k}\|_2^2 + \\
    \frac{\alpha }{2}(\|e_k^* - {e_{k - 1}}\|_2^2 - \|{e_*} - {e_k}\|_2^2) \\
    -  \left\langle \frac{{{P_k}}}{\eta }({p_k} - {p_{k - 1}}),{p_k} - p_k^*\right\rangle 
    \end{gathered}
\end{equation}
where $\triangledown \|p\|_k$ denotes the gradience.\\
\textbf{Proof}: from (8a), one can obtain\\
\begin{equation}\label{E20}
  \begin{gathered}
   - \frac{{{P_k}}}{\eta }({p_k} - {p_{k - 1}}) \\
   - {{\cal A}}_k^{\dagger}( - {\lambda _{k - 1}} + \alpha ({{{\cal A}}_k}{p_k} + {e_{k - 1}} - {b_k})) = \triangledown \|{p_k}\|_2
  \end{gathered}
\end{equation}
from (8b), one can obtain\\
\begin{equation}\label{E21}
  \begin{gathered}
    \frac{1}{\gamma }{e_k} + \alpha ({{{\cal A}}_k}{p_k} + {e_k} - {b_k}) - {\lambda _{k - 1}} = 0\\
    \frac{1}{\gamma }{e_k} =  - {\lambda _k}
    \end{gathered}
\end{equation}
by using \eqref{E20},\eqref{E21}, it holds that
\begin{equation}\label{E22}
  \begin{gathered}
    \|{p_k}\|{_2} - \|p_k^*\|{_2} \\
    \le  -  \left\langle {{\cal A}}_k^{\dagger}( - {\lambda _k} + \alpha ({e_k} - {e_{k - 1}})),{p_k} - p_k^*\right\rangle\\
     -  \left\langle \frac{{{P_k}}}{\eta }({p_k} - {p_{k - 1}}),{p_k} - p_k^*\right\rangle \\
     =  -  \left\langle - {\lambda _k} + \alpha ({e_{k - 1}} - {e_k}),{{{\cal A}}_k}({p_k} - p_k^*)\right\rangle\\
     -  \left\langle \frac{{{P_k}}}{\eta }({p_k} - {p_{k - 1}}),{p_k} - p_k^*\right\rangle\\
     =  -  \left\langle - {\lambda _k},{{{\cal A}}_k}{p_k} - {b_k} + e_k^*\right\rangle\\
     + \alpha  \left\langle {e_k} - {e_{k - 1}},{{{\cal A}}_k}{p_k} - {b_k} + e_k^*\right\rangle\\
     -  \left\langle \frac{{{P_k}}}{\eta }({p_k} - {p_{k - 1}}),{p_k} - p_k^*\right\rangle\\
     =  -  \left\langle - {\lambda _k},{{{\cal A}}_k}{p_k} - {b_k} + e_k^*\right\rangle\\
     + \frac{\alpha }{2}(\|e_k^* - {e_{k - 1}}\|_2^2 + \|e_k^* - {e_k}\|_2^2)\\
     + \frac{\alpha }{2}{\|{{\cal A}}_k}{p_k} + {e_k} - {b_k}\|_2^2 - \frac{\alpha }{2}{\|{{\cal A}}_k}{p_k} + {e_{k - 1}} - {b_k}\|_2^2
    \end{gathered}
\end{equation}
The last step comes from the formula\\
\begin{equation}\label{E23}
  \begin{gathered}
    \left\langle {u_1} - {u_2},{u_3} + {u_4}\right\rangle = \\
    \frac{1}{2}(\|{u_4} - {u_2}\|_2^2 - \|{u_4} - {u_1}\|_2^2 \\
    + \|{u_3} + {u_1}\|_2^2 - \|{u_3} + {u_2}\|_2^2)
  \end{gathered}
\end{equation}
the same can be obtained by (8b)
\begin{equation}\label{E24}
  \begin{gathered}
    \frac{1}{{2\gamma }}\|{e_k}\|_2^2 - \frac{1}{{2\gamma }}\|e_k^*\|_2^2 \le  -  \left\langle - {\lambda _k},{e_k} - e_k^*\right\rangle
  \end{gathered}
\end{equation}
add formulas \eqref{E22} and \eqref{E24}, one can obtain
\begin{equation}\label{E25}
  \begin{gathered}
      \|{p_k}\|{_2} + \frac{1}{{2\gamma }}\|{e_k}\|_2^2 - (\|p_k^*\|{_2} + \frac{1}{{2\gamma }}\|e_k^*\|_2^2)\\
       \le  -  \left\langle - {\lambda _k},{{{\cal A}}_k}{p_k} + {e_k} - {b_k}\right\rangle \\
       + \frac{\alpha }{2}{\|{{\cal A}}_k}{p_k} + {e_k} - {b_k}\|_2^2\\
       + \frac{\alpha }{2}(\|e_k^* - {e_{k - 1}}\|_2^2 + \|e_k^* - {e_k}\|_2^2) \\
       - \frac{\alpha }{2}{\|{{\cal A}}_k}{p_k} + {e_{k - 1}} - {b_k}\|_2^2
  \end{gathered}
\end{equation}
based on (8c), the first two terms of formula \eqref{E25} can be rewritten as
\begin{equation}\label{E26}
  \begin{gathered}
      -  \left\langle - {\lambda _k},{{{\cal A}}_k}{p_k} + {e_k} - {b_k}\right\rangle + \frac{\alpha }{2}{\|{{\cal A}}_k}{p_k} + {e_k} - {b_k}\|_2^2\\
      = \frac{1}{{2\alpha }}(2 \left\langle - {\lambda _k},{\lambda _k} - {\lambda _{k - 1}}\right\rangle + \|{\lambda _k} - {\lambda _{k - 1}}\|_2^2)\\
      = \frac{1}{{2\alpha }}(\|{\lambda _{k - 1}}\|_2^2 - \|{\lambda _k}\|_2^2)
  \end{gathered}
\end{equation}
Bring \eqref{E26} into \eqref{E25}, we can obtain the result \eqref{E19}.\\
The proof is complete.
\par Theorem 1 is proved as follows:\\
\textbf{Proof:} from lemma 1, The last item of \eqref{E19} can be rewritten as
\begin{equation}\label{E27}
  \begin{gathered}
      -  \left\langle \frac{{{P_k}}}{\eta }({\rho _k} - {\rho _{k - 1}}),{p_k} - p_k^*\right\rangle  = \\
      \frac{1}{{2\eta }}\|p_{k - 1}^* - {p_{k - 1}}\|_{{P_k}}^2 - \frac{1}{{2\eta }}\|p_k^* - {p_k}\|_{{P_k}}^2\\
     {\rm{                                             }} - \frac{1}{{2\eta }}\|{p_k} - {p_{k - 1}}\|_{{P_k}}^2
  \end{gathered}
\end{equation}
based on the parameter condition \eqref{E4}, $\frac{1}{2}\left\| p \right\|_{{P_k}}^2$ is a strongly convex function. $\exists \beta  > 0$, ${P_k} - \beta I > 0$. By the FenChel-Young inequality, one has:
\begin{equation}\label{E28}
  \begin{gathered}
      \|{p_{k - 1}}\|{_2} - \|{p_k}\|{_2} \le \; \left\langle \bigtriangledown \|{p_{k - 1}}\|_2,{p_{k - 1}} - {p_k}\right\rangle\\
       = \left\langle 2p_{k-1},{p_{k - 1}} - {p_k}\right\rangle\\ 
       =  \left\langle 2\sqrt {\frac{\eta }{\beta }} \|p_{k-1}\|_2^2,\sqrt {\frac{\beta }{\eta }} ({p_{k - 1}} - {p_k})\right\rangle\\
       \le \frac{\eta }{{\beta }}\|\|{p_{k - 1}}\|_2^{'}\|_2^2 + \frac{\beta }{{\eta }}\|{p_{k - 1}} - {p_k}\|_2^2
  \end{gathered}
\end{equation}
hence, \eqref{E19} can be written as
\begin{equation}\label{E29}
  \begin{gathered}
      \|{p_k}\|{_2} + \frac{1}{{2\gamma }}\|{e_k}\|_2^2 - (\|p_k^*\|{_2} + \frac{1}{{2\gamma }}\|e_k^*\|_2^2)\\
       \le \frac{1}{{2\alpha }}(\|{\lambda _{k - 1}}\|_2^2 - \|{\lambda _k}\|_2^2) - \frac{\alpha }{2}{\|{{\cal A}}_k}{p_k} + {e_{k - 1}} - {b_k}\|_2^2\\
       + \frac{\alpha }{2}(\|e_k^* - {e_{k - 1}}\|_2^2 - \|e_k^* - {e_k}\|_2^2) + \\
      \frac{\eta}{{\beta}}\|p_{k-1}\|_2^2 + \frac{1}{{\eta}}(\|p_k^* - {p_{k - 1}}\|_{P_k}^2 - \|p_k^* - {p_k}\|_{P_k}^2)
  \end{gathered}
\end{equation}
Since $\left\{ {p_k^*,e_k^*} \right\}$ are finite variable matrix in the tracking process, $\exists {G_p} > 0$, $\|p\|_k\leq G_p$. $\left\| {p_1^* - {p_1}} \right\|_{{P_k}}^2,\left\| {e_1^* - {e_1}} \right\|_{{P_k}}^2$ are bounded, for any $\left\{ {p_k^*,e_k^*} \right\}$ satisfied ${{\cal A}_k}p_k^* + e_k^* = {b_k}$, we have $\left\| {p_k^* - {p_1}} \right\|_{{P_k}}^2 \le D_p^2,\left\| {e_k^* - {e_1}} \right\|_{{P_k}}^2 \le {D_e}$. For $\lambda_1$, we set $\lambda_1=0$. Using the conditions \eqref{E4}, the following holds
\begin{equation}\label{E30}
  \begin{gathered}
      {R_1}(T) = \frac{1}{T}(\sum\limits_{k = 1}^T {\|{p_{k - 1}}\|{_2}} + \frac{1}{{2\gamma }}\|e\|_2^2 - (\|p_k^*\|{_2} + \frac{1}{{2\gamma }}\|e_k^*\|_2^2) \\
      + \frac{1}{{2\gamma }}\|{e_1}\|_2^2 - \frac{1}{{2\gamma }}\|{e_T}\|_2^2)\\
       \le \frac{1}{T}(\frac{1}{{2\alpha }}(\|{\lambda _1}\|_2^2 - \|{\lambda _T}\|_2^2) + \frac{\alpha }{2}(\|e_1^* - {e_1}\|_2^2 - \|e_T^* - {e_T}\|_2^2))\\
       + \frac{\eta }{{\beta }}\sum\limits_{k = 1}^T\|p_{k-1}\|_2^2 + \frac{1}{{\eta }}(\|p_1^* - {p_1}\|_{{P_k}}^2 - \|p_T^* - {p_T}\|_{{P_k}}^2)\\
       \le \frac{\alpha }{{2T}}D_e^2 + \frac{{D_p^2}}{{\eta T}} + \frac{{\eta G_p^2}}{{\beta}}\\
       =\frac{D_e^2}{2\sqrt{T}}+\frac{\sigma_m D_p^2}{\tau \sqrt{T}}+\frac{\tau G_p^2}{\beta \sigma_m \sqrt{T}}=O(\frac{1}{\sqrt{T}})
  \end{gathered}
\end{equation}
\par In the tracking process, the distance between the reconstructed value and the optimal value theoretically decreases as a whole. So $\exists k > 0$ make $\left| {{{\left\| {{p_k}} \right\|}_2} + \frac{1}{{2\gamma }}\left\| {{e_k}} \right\|_2^2 - ({{\left\| {p_k^*} \right\|}_2} + \frac{1}{{2\gamma }}\left\| {e_k^*} \right\|_2^2)} \right|$ get the maximum. we set $F>$ maximum, for any $\left\{ {{{\hat p}_*},{{\hat e}_*}} \right\}$, $\forall k,\exists F > 0,{\left\| {p_k^*} \right\|_2} + \frac{1}{{2\gamma }}\left\| {e_k^*} \right\|_2^2 - ({\left\| {p_k^*} \right\|_2} + \frac{1}{{2\gamma }}\left\| {e_k^*} \right\|_2^2) \ge  - F$. Reorder formula \eqref{E26}:
\begin{equation}\label{E31}
  \begin{gathered}
      {\|{{\cal A}}_k}{p_k} + {e_{k - 1}} - {b_k}\|_2^2 \le \frac{{2F}}{\alpha } \\
      + \frac{1}{{{\alpha ^2}}}(\|{\lambda _{k - 1}}\|_2^2 - \|{\lambda _k}\|_2^2) \\
      + (\|e_k^* - {e_{k - 1}}\|_2^2 - \|e_k^* - {e_k}\|_2^2)\\
       + \frac{1}{{\alpha \eta }}(\|p_{k - 1}^* - {p_{k - 1}}\|_{{P_k}}^2 \\
       - \|p_k^* - {p_k}\|_{{P_k}}^2 - \|{p_k} - {p_{k - 1}}\|_{{P_k}}^2)
  \end{gathered}
\end{equation}
\par Thus it holds that
\begin{equation}\label{E32}
  \begin{gathered}
      {R_2}(T) = \frac{1}{T}(\sum\limits_{k = 1}^T {{\|{{\cal A}}_k}{p_k} + {e_k} - {b_k}\|_2^2} ) \le \\
      \frac{1}{T}(\sum\limits_{k = 1}^T {{\|{{\cal A}}_k}{p_k} + {e_{k - 1}} - {b_k}\|_2^2} ) \le \\
      \frac{{2F}}{\alpha } + \frac{1}{{{\alpha ^2}T}}(\|{\lambda _1}\|_2^2 - \|{\lambda _T}\|_2^2) \\
      + (\|{e_*} - {e_1}\|_2^2 - \|{e_*} - {e_T}\|_2^2)\\
       + \frac{1}{{\alpha \eta T}}(\|p_1^* - {p_1}\|_{{P_k}}^2 - \|p_T^* - {p_T}\|_{{P_k}}^2)\\
       \le \frac{{2F}}{\alpha } + \frac{{D_e^2}}{T} + \frac{2}{{\alpha \eta T}}D_p^2\\
       = \frac{2F}{\sqrt{T}}+\frac{D_e^2}{T}+\frac{2\sigma_mD_p^2}{\beta \tau T}=O(\frac{1}{\sqrt{T}})
  \end{gathered}
\end{equation}
Based on the result of \eqref{E30} and \eqref{E32}, we can obtain the \eqref{E10}.\\
The proof is complete.
\par The results of theorem 1 show that the convergence order of boundary of loss function $R_1(T)$ and $R_2(T)$ is $O(\frac{1}{\sqrt{T}})$. With the increase of the times $T$, the two loss functions decrease with the order greater than $O(\frac{1}{\sqrt{T}})$; This explains that with the increase of the sampling times $T$, the distance between the value of reconstructed variables and the optimal value gradually decreases, indicating that the algorithm has convergence and the convergence value can meet the constraints of the optimization problem.\\
\subsection{Proof of Theorem 2}
\par Theorem 2 is proved as follows:\\
\par Conclusions from Theorem 1 \eqref{E10}:
\begin{equation}\label{E33}
  \begin{gathered}
    {R_1}(T) = \frac{1}{T}(\sum\limits_{k = 1}^T {({{\left\| {{p_k}} \right\|}_2} + \frac{1}{{2\gamma }}\left\| {{e_k}} \right\|_2^2)}  - \\
    \sum\limits_{k = 1}^T {({{\left\| {p_k^*} \right\|}_2} + \frac{1}{{2\gamma }}\left\| {e_k^*} \right\|_2^2)} ) \le O(\frac{1}{{\sqrt T }})
  \end{gathered}
\end{equation}
\begin{equation}\label{E34}
  \begin{gathered}
    {R_2}(T) = \frac{1}{T}(\sum\limits_{k = 1}^T {\left\| {{{\cal A}_k}{p_k} + {e_k} - {b_k}} \right\|_2^2)}  \le O(\frac{1}{{\sqrt T }})
  \end{gathered}
\end{equation}
by Jensen inequality:
\begin{equation}\label{E35}
  \begin{gathered}
    ({\left\| {{p_T}} \right\|_2} + \frac{1}{{2\gamma }}\left\| {{e_T}} \right\|_2^2) - ({\left\| {p_k^*} \right\|_2} + \frac{1}{{2\gamma }}\left\| {e_k^*} \right\|_2^2) \\
    \le O(\frac{1}{{\sqrt T }})
  \end{gathered}
\end{equation}
\begin{equation}\label{E36}
  \begin{gathered}
    \left\| {{{\cal A}_T}{p_T} + {e_T} - {b_T}} \right\|_2^2 \le O(\frac{1}{{\sqrt T }})
  \end{gathered}
\end{equation}
from \eqref{E36}, we obtain
\begin{equation}\label{E37}
  \begin{gathered}
      \left\| {{{\cal A}_k}({p_k} - p_T^*)} \right\|_2^2 \le \left\| {{{\cal A}_k}} \right\|_2^2\left\| {{p_k} - p_T^*} \right\|_2^2\\
       \le \left\| {{{\cal A}_k}} \right\|_2^2\left\| {({b_k} - {e_k}) - ({b_k} - e_k^*)} \right\|_2^2 + O(\frac{1}{{\sqrt T }})\\
       \le \left\| {{{\cal A}_k}} \right\|_2^2\left\| {e_k^* - {e_k}} \right\|_2^2 + O(\frac{1}{{\sqrt T }})\\
       \le O(\frac{1}{{\sqrt T }})
  \end{gathered}
\end{equation}
divide both sides of equation \eqref{E37} by $\left\| {{{\cal A}_k}} \right\|_2^2$ to get
\begin{equation}\label{E38}
  \begin{gathered}
    \left\| {{p_T} - p_T^*} \right\|_2^2 \le O(\frac{1}{{\sqrt T }})
  \end{gathered}
\end{equation}
Equation \eqref{E38} shows that with the increase of sampling times $T$, the normalized distance of the reconstruction error converges to the order $O(\frac{1}{{\sqrt T }})$.
\begin{equation}\label{E39}
  \begin{gathered}
    D(\rho _T^*,{\rho _T}) = D(p_T^*,{p_T}) \le O(\frac{1}{{\sqrt T }})
  \end{gathered}
\end{equation}
The proof is complete.
\par The results of theorem 2 show that the convergence order of the normalized distance boundary is $O(\frac{1}{\sqrt{T}})$. This shows that as the times $T$ increases, the reconstructed value gradually converges to the optimal value with the velocity greater than $O(\frac{1}{\sqrt{T}})$.
\subsection{Proof of theorem3}

The proof process is as follows: Firstly, it is proved by Lemma 2 that the $l_2$ norm distance of the value of  reconstructed variables and the optimal value decreases at a speed less than $\beta$ times at sampling times $k+1$. Lemma 3 is used to prove that the cumulative distance of optimization function of the value of reconstructed variables and optimal value at times $T$ is bounded. Lemma 4 is used to prove the average value of the gradient $\nabla {h_k}({p_k})$ at $T$ times is bounded. Then, we use the conclusion of three lemmas to prove the convergence order of loss function $REG(T)$ in Theorem 3. Finally, theorem 4 is proved by the conclusion of theorem 3.\\
\textbf{Lemma 2:} for KF-QSE algorithm, we have
\begin{equation}\label{E40}
  \begin{gathered}
    \|{p_k} - p_{k - 1}^*\|{_2} \le \beta \|{p_{k - 1}} - p_{k - 1}^*\|{_2} + {\left\| {{E_{k - 1}}} \right\|_2}
  \end{gathered}
\end{equation}
where $\beta = 1- \mu$.\\
\textbf{Proof:}
we define
\begin{equation}\label{E41}
  \begin{gathered}
    pro{x_{{g_k}}}(x) = \arg\mathop {\min }\limits_u {g_k}(u) + \|u - x\|_2^2\\
    {g_k}(p) = \frac{1}{{2\gamma }}\|{\mathcal{A}_k}p - {b_k}\|_2^2
  \end{gathered}
\end{equation}
for $h_k(p)$ and $g_k(p)$, both have the same gradient at $p_{k-1}$, we can rewrite \eqref{E13} as 
\begin{equation}\label{E42}
  {\hat p_k} = {p_{k - 1}} - \nabla {h_k}({p_{k - 1}}) = {p_{k - 1}} - \nabla {g_k}({p_{k - 1}})
\end{equation}
hence, we rewrite the \eqref{E42}
\begin{equation}\label{E43}
  {p_k} = {p_{k - 1}} - \tilde \nabla {g_k}({p_{k - 1}}) = {p_{k - 1}} - \nabla {g_k}({p_{k - 1}}) + {E_k}
\end{equation}
where ${E_k} = {p_k} - {\hat p_k}$.\\
under definition \eqref{E41}, \eqref{E16}can be rewritten as ${p_k} = pro{x_{{g_k}}}({p_{k - 1}})$. By Peter-Paul inequality
\begin{equation}\label{E44}
  \begin{gathered}
      p_k^* = pro{x_{{g_{k - 1}}}}(p_k^*)\\
      \|pro{x_{{g_{k - 1}}}}(u) - pro{x_{{g_{k - 1}}}}(v)\|{_2} \le \|u - v\|{_2}
  \end{gathered}
\end{equation}
from defination \eqref{E41}, we know that ${g_k}(p)$ is a $\beta$-convex function, $\exists \mu>0$, it holds
\begin{equation}\label{E45}
  \begin{gathered}
    < \nabla {g_k}(x) - \nabla {g_k}(y),x - y >  \ge \mu {\left\| {x - y} \right\|^2},{\rm{  }}\forall {\rm{x,y > 0}}
  \end{gathered}
\end{equation}
let ${L_g} \ge \max \{ {\left\| {{{\cal A}_k}} \right\|_2}\} ,k = 1, \ldots ,T$, we can obtain that ${g_k}(p)$ is $L_g$-Lipschitz continuous in $R$
\begin{equation}\label{E46}
  \begin{gathered}
    \left\| {\nabla {g_k}(x) - \nabla {g_k}(y)} \right\| \le {L_g}\left\| {x - y} \right\|,{\rm{  }}\forall {\rm{x,y > 0}}
  \end{gathered}
\end{equation}
Since $g_k(x)$ is $L_g$-continuous, let
\begin{equation}\label{E47}
  \begin{gathered}
    g_k^c(x) = {g_k}(x) - \frac{\mu }{2}\|x\|_2^2
  \end{gathered}
\end{equation}
based on \eqref{E47}, it holds that $g_k^c(x)$ is $L_g$-continuous. Now we have
\begin{equation}\label{E48}
  \begin{gathered}
      {(\nabla {g_k}(x) - \nabla {g_k}(y))^T}(x - y) \ge \frac{{\mu {L_g}}}{{{L_g} + \mu }}\|x - y\|_2^2\\
       + \frac{1}{{{L_g} + \mu }}\|\nabla {g_k}(x) - \nabla {g_k}(y)\|_2^2
  \end{gathered}
\end{equation}
from \eqref{E44}, we obtain
\begin{equation}\label{E49}
  \begin{gathered}
      \|{p_k} - p_{k - 1}^*\|_2^2 = \|pro{x_{{g_{k - 1}}}}({p_{k - 1}}) - pro{x_{{g_{k - 1}}}}(p_{k - 1}^*)\|_2^2\\
       = \|({p_{k - 1}} - \tilde \nabla {g_{k - 1}}({p_{k - 1}})) - (p_{k - 1}^* - \nabla {g_{k - 1}}(p_{k - 1}^*))\|_2^2\\
       = \|({p_{k - 1}} - \nabla {g_{k - 1}}({p_{k - 1}})) - (p_{k - 1}^* - \nabla {g_{k - 1}}(p_{k - 1}^*)) - {E_{k - 1}}\|_2^2\\
       = \|({p_{k - 1}} - \nabla {g_{k - 1}}({p_{k - 1}})) - (p_{k - 1}^* - \nabla {g_{k - 1}}(p_{k - 1}^*))\|_2^2\\
       + \left\| {{E_{k - 1}}} \right\|_2^2 - 2E_{k - 1}^T[({p_{k - 1}} \\
       - \nabla {g_{k - 1}}({p_{k - 1}})) - (p_{k - 1}^* - \nabla {g_{k - 1}}(p_{k - 1}^*))]
  \end{gathered}
\end{equation}
the first item in \eqref{E49} can be rewritten as 
\begin{equation}\label{E50}
  \begin{gathered}
      \|({p_{k - 1}} - \nabla {g_{k - 1}}({p_{k - 1}})) - (p_{k - 1}^* - \nabla {g_{k - 1}}(p_{k - 1}^*))\|_2^2\\
       = \|{p_{k - 1}} - p_{k - 1}^*\|_2^2 + \|\nabla {g_{k - 1}}({p_{k - 1}}) - \nabla {g_{k - 1}}(p_{k - 1}^*)\|_2^2\\
       - 2{(\nabla {g_{k - 1}}({p_{k - 1}}) - \nabla {g_{k - 1}}(p_{k - 1}^*))^T}({p_{k - 1}} - p_{k - 1}^*)\\
       = \|{p_{k - 1}} - p_{k - 1}^*\|_2^2 + \|\nabla {g_{k - 1}}({p_{k - 1}}) - \nabla {g_{k - 1}}(p_{k - 1}^*)\|_2^2\\
       - 2(\frac{{\mu {L_g}}}{{{L_g} + \mu }}\|{p_{k - 1}} - p_{k - 1}^*\|_2^2 + \frac{1}{{{L_g} + \mu }}\|\nabla {g_{k - 1}}({p_{k - 1}}) - \\
       \nabla {g_{k - 1}}(p_{k - 1}^*)\|_2^2)\\
       \le (1 - \frac{{2{L_g}\mu }}{{{L_g} + \mu }})\|{p_{k - 1}} - p_{k - 1}^*\|_{2}^2\\
       + (1 - \frac{2}{{{L_g} + \mu }})\|\nabla {g_{k - 1}}({p_{k - 1}}) - \nabla {g_{k - 1}}(p_{k - 1}^*)\|_2^2\\
       \le (1 - \frac{{2{L_g}\mu }}{{{L_g} + \mu }})\|{p_{k - 1}} - p_{k - 1}^*\|_2^2 + {\mu ^2}(1 \\
       - \frac{2}{{{L_g} + \mu }})\|{p_{k - 1}} - p_{k - 1}^*\|_2^2\\
       = {\beta ^2}\|{p_{k - 1}} - p_{k - 1}^*\|_2^2
  \end{gathered}
\end{equation}
define $Er{r_T} = \frac{D_{\rho}}{{T(1 - \beta )}}\sum\limits_{k = 1}^T {{{\left\| {{E_k}} \right\|}_2}}$. Bring \eqref{E50} results back to \eqref{E48}, we have
\begin{equation}\label{E51}
  \begin{gathered}
      \|{p_k} - p_{k - 1}^*\|_2^2 = \|pro{x_{{g_{k - 1}}}}({p_{k - 1}}) - pro{x_{{g_{k - 1}}}}(p_{k - 1}^*)\|_2^2\\
       \le {\beta ^2}\|{p_{k - 1}} - p_{k - 1}^*\|_2^2 + \left\| {{E_{k - 1}}} \right\|_2^2 \\+ 2\beta {\left\| {{E_{k - 1}}} \right\|_2}\|{p_{k - 1}} - p_{k - 1}^*\|{_2}\\
  \end{gathered}
\end{equation}
from \eqref{E51}, we know that \eqref{E40} holds.\\
The proof is complete.\\
\textbf{Lemma 3:} for KF-QSE algorithm, we have
\begin{equation}\label{E52}
  \begin{gathered}
    \sum\limits_{k = 1}^T {\|{p_k} - p_k^*\|{_2}}  \le \frac{1}{{1 - \beta }}(\|{p_1} - p_1^*\|{_2} + {W_T} + \sum\limits_{k = 1}^T {{{\left\| {{e_k}} \right\|}_2}} )
  \end{gathered}
\end{equation}
where $W_T=\sum_{k=2}^{T}\|p_k^*-p_{k-1}^*\|_2^2$.\\
\textbf{Proof:} By the triangle inequality
\begin{equation}\label{E53}
  \begin{gathered}
      \sum\limits_{k = 2}^T {\|{p_k} - p_k^*\|{_2}}  = \sum\limits_{k = 2}^T {\|{p_k} - p_{k - 1}^* + p_{k - 1}^* - p_k^*\|{_2}} \\
       \le \sum\limits_{k = 1}^T \beta  \|{p_k} - p_k^*\|{_2} + \sum\limits_{k = 1}^T {{{\left\| {{e_k}} \right\|}_2}}  + {W_T}
  \end{gathered}
\end{equation}
Add $\|{p_1} - p_1^*\|{_2}$ to both sides of \eqref{E53}
\begin{equation}\label{E54}
  \begin{gathered}
      \sum\limits_{k = 1}^T {\|{p_k} - p_k^*\|{_2}}  \le \|{p_1} - p_1^*\|{_2} + \sum\limits_{k = 1}^T \beta  \|{p_k} - p_k^*\|{_2} + {W_T}\\
       = \frac{1}{{1 - \beta }}\|{p_1} - p_1^*\|{_2} + \frac{1}{{1 - \beta }}\sum\limits_{k = 1}^T {{{\left\| {{e_k}} \right\|}_2}}  + \frac{1}{{1 - \beta }}{W_T}
  \end{gathered}
\end{equation}
The proof is complete.\\
\textbf{Lemma 4:} the function $\nabla {h_k}({\rho _k})$ satisfies
\begin{equation}\label{E55}
  \begin{gathered}
    \|\nabla {h_k}({\rho _k})\|{_2} \le D: = (1 + \beta )\|{p_1} - p_1^*\|{_2} + \frac{\sigma(1+\beta) }{{1 - \beta }}
  \end{gathered}
\end{equation}
\textbf{Proof:} from \eqref{E41}, we have
\begin{equation}\label{E56}
  \begin{gathered}
    {p_{k - 1}} - {p_k} = \tilde \nabla {g_k}({p_k})
  \end{gathered}
\end{equation}
let
\begin{equation}\label{E57}
  \begin{gathered}
    {G_k}({p_k}) = \tilde \nabla {g_k}({p_k})
  \end{gathered}
\end{equation}
we obtain
\begin{equation}\label{E58}
  \begin{gathered}
    {G_k}({p_k}) = ({p_{k - 1}} - p_{k - 1}^*) - ({p_k} - p_{k - 1}^*)
  \end{gathered}
\end{equation}
by using the triangle inequality, $G_k(p_k)$ holds
\begin{equation}\label{E59}
  \begin{gathered}
    \|{G_k}({p_k})\|{_2} \le \|{p_{k - 1}} - p_{k - 1}^*\|{_2} - \\
    \|{p_k} - p_{k - 1}^*\|{_2} \le (1 + \beta )\|{p_{k - 1}} - p_{k - 1}^*\|{_2}
  \end{gathered}
\end{equation}
based on \eqref{E59}, we have 
\begin{equation}\label{E60}
  \begin{gathered}
    \|\nabla {h_k}({\rho _k})\|{_2} = \|\nabla {g_k}({p_k})\|{_2} = \\
    \|{G_k}({p_k})\|{_2} \le (1 + \beta )\|{p_{k - 1}} - p_{k - 1}^*\|{_2}
  \end{gathered}
\end{equation}
we set $\sigma$ as the supremacy of set $\{ \left\| {p_k^* - p_{k - 1}^*} \right\|\} ,k = 1, \ldots ,T$, based on \eqref{E40}, we have
\begin{equation}\label{E61}
  \begin{gathered}
      \|{p_k} - p_k^*\|{_2} = \|{p_k} - p_{k - 1}^* + p_{k - 1}^* - p_k^*\|{_2}\\
       \le \|{p_k} - p_{k - 1}^*\|{_2} + \|p_{k - 1}^* - p_k^*\|{_2}\\
       \le \beta \|{p_{k - 1}} - p_{k - 1}^*\|{_2} + \sigma \\
       \le {\beta ^{k - 1}}\|{p_1} - p_1^*\|{_2} + \sigma (\frac{{1 - {\beta ^{k - 1}}}}{{1 - \beta }})\\
       \le \|{p_1} - p_1^*\|_2 + (\frac{\sigma }{{1 - \beta }})
  \end{gathered}
\end{equation}
based on \eqref{E61}, it holds
\begin{equation}\label{E62}
  \begin{gathered}
    \|\nabla {h_k}({p_k})\|{_2} \le \|{p_1} - p_1^*\|{_2} + \frac{\sigma(1+\beta) }{{1 - \beta }}
  \end{gathered}
\end{equation}
The proof is complete.
\par The proof of Theorem 3 is as follows:\\
\textbf{Proof:} 
\begin{equation}\label{E63}
  \begin{gathered}
      \mathop \sum \limits_{k = 1}^T [{h_k}({p_k}) - {h_k}(p_k^*)] \le \sum\limits_{k = 1}^T  \left\langle \nabla {h_k}({p_k}),{p_k} - p_k^*\right\rangle\\
       \le \sum\limits_{k = 1}^T {\|\nabla {h_k}(} {p_k})\|{_2} \cdot \|{p_k} - p_k^*\|{_2}\\
       \le \sum\limits_{k = 1}^T D  \cdot \|{p_k} - p_k^*\|{_2}\\
       \le \frac{D}{{1 - \beta }}\|{p_1} - p_1^*\|{_2} + \frac{D}{{1 - \beta }}{W_T} + \frac{D}{{1 - \beta }}\sum\limits_{k = 1}^T {{{\left\| {{e_k}} \right\|}_2}} 
  \end{gathered}
\end{equation}
By definition \eqref{E15}, we have
\begin{equation}\label{E64}
  \begin{gathered}
    REG(T) \le \frac{D}{{T(1 - \beta )}}\|{p_1} - p_1^*\|{_2} + \frac{D}{{T(1 - \beta )}}{W_T} \\
    + \frac{D}{{T(1 - \beta )}}\sum\limits_{k = 1}^T {{{\left\| {{e_k}} \right\|}_2}}
  \end{gathered}
\end{equation}
\par From Theorem 3, we obtain that the loss function $REG(T)$ converges to range $\frac{D}{{T(1 - \beta )}}\sum\limits_{k = 1}^T {{{\left\| {{E_k}} \right\|}_2}}$ with a convergence order of ${\cal O}(\frac{1}{T})$.
\subsection{The proof of theorem 4}
\par The proof of Theorem 4 is as follows:\\
From Theorem 3, conclusion \eqref{E17}
\begin{equation}\label{E65}
  \begin{gathered}
      REG(T) = \frac{1}{T}\sum\limits_{k = 1}^T {{{\left\| {{p_k} - p_k^*} \right\|}_2}} \\
       \le \frac{D}{{T(1 - \beta )}}{\left\| {{p_1} - p_1^*} \right\|_2} + \frac{D}{{T(1 - \beta )}}{W_k} \\
       + \frac{D}{{T(1 - \beta )}}\sum\limits_{k = 1}^T {{{\left\| {{E_k}} \right\|}_2}} 
  \end{gathered}
\end{equation}
since the difference between the reconstructed value and the optimal value continues to decrease as tracking proceeds, there is
\begin{equation}\label{E66}
  \begin{gathered}
    {\left\| {{p_T} - p_T^*} \right\|_2} \le \frac{D}{{(1 - \beta )T}}{\left\| {{p_1} - p_1^*} \right\|_2} \\
    + \frac{D}{{(1 - \beta )T}}{W_T} + \frac{D}{{T(1 - \beta )}}\sum\limits_{k = 1}^T {{{\left\| {{E_k}} \right\|}_2}}
  \end{gathered}
\end{equation}
by using Jensen's inequality
\begin{equation}\label{E67}
  \begin{gathered}
      {\rm{                 }}D(\rho _T^*,{\rho _T}) = D(p_T^*,{p_T}) \le {\left\| {{p_T} - p_T^*} \right\|_2}\\
       \le \frac{D}{{(1 - \beta )T}}{\left\| {{p_1} - p_1^*} \right\|_2} + \frac{D}{{(1 - \beta )T}}{W_T} \\
       + \frac{D}{{T(1 - \beta )}}\sum\limits_{k = 1}^T {{{\left\| {{E_k}} \right\|}_2}} 
  \end{gathered}
\end{equation}
\par Equation \eqref{E67} illustrates that as the number of sampling times $T$ of the KF-QSE algorithm increases, the normalized distance $D(p_T^*,{p_T})$ converges to the range of the error term $Er{r_T}$ at the order of $O(\frac{1}{T})$.
\begin{equation}\label{E68}
  \begin{gathered}
    D(p_T^*,{p_T}) \le O(\frac{1}{T}) + Er{r_T}
  \end{gathered}
\end{equation}
The proof is complete.
\par In summary, from the conclusions of Theorem 2 and Theorem 4, we know that the KF-QSE has a faster convergence rate than the OPG-ADMM in terms of real-time tracking of quantum states, but the normalized distance of KF-QSE will eventually decrease because $Err_T$ does not decrease with the number of sampling times $T$. The influence of the $Err_T$ makes the normalized distance $D(p_T^*,{p_T})$ of KF-QSE unable to be reduced within a certain range. The normalized distance of the OPG-ADMM can be continuously reduced as the times of sampling increases.
\section{Numerical Simulation experiment}
According to theorem 2 and theorem 4, the normalized distance converges to 0 at the rate $O(\frac{1}{\sqrt{T}})$ by using OPG-ADMM algorithm. But for the KF-QSE, the convergence term of the normalized distance contains the error term $Err_T$ caused by the gradient error, this makes the KF-QSE algorithm have a higher convergence rate $O(\frac{1}{T})$ but the normalized distance cannot converge to 0. Therefore, if the sampling times $T$ is large, the theoretical reconstruction accuracy of KF-QSE will be smaller than the OPG-ADMM. To verify this, we perform experiments under fixed sliding window length to compare the algorithm performance.
\par In Numerical experiments, We compare the normalized distance of quantum density matrix and sampling times of KF-QSE and OPG-ADMM algorithms for 4-qubit dynamic quantum states in fixed $l=40$. Figure 1 shows the normalized distance of the online reconstruction state of the two algorithms at each sampling times.
\begin{figure}
  \centering
  \includegraphics[width=3.5in]{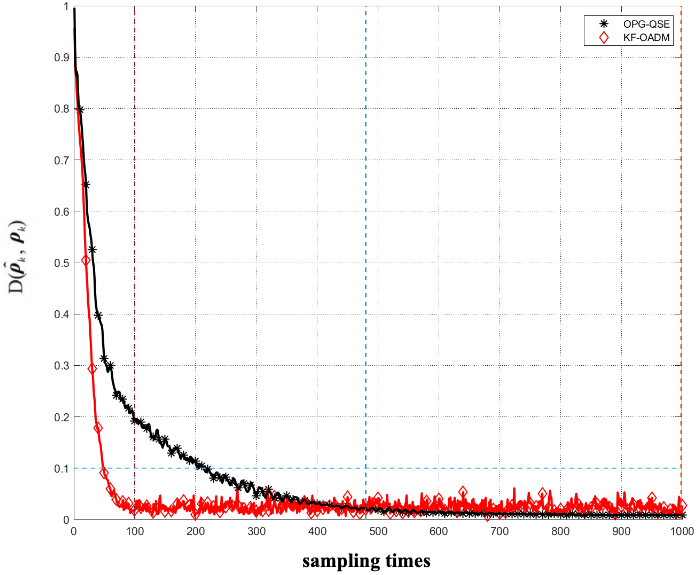}
  \caption{Numerical performance comparison results  of  4-qubit system}
  \label{fig_1}
\end{figure}
\par From Figure 1, the normalized distance of both KF-QSE and OPG-ADMM can gradually decrease, indicating that both algorithms can achieve certain accuracy of quantum state reconstruction. When the sampling time $k$ is small, the $Err_T$ in KF-QSE has little influence, and the convergence rate of KF-QSE is obviously higher than that of OPG-ADMM. For the normalized distance accuracy of 0.1, KF algorithm can reach the accuracy when the sampling time reaches 48 times, and OPG can reach the accuracy only when the sampling time reaches 223 times. This state of KF-QSE continues until the sampling time reaches about $k=100$. For the reconstruction state at $k=100$, the normalized distance of KF-QSE and OPG-ADMM algorithms are 0.090 and 0.127 respectively, which reflects the difference in convergence order between the $O(\frac{1}{\sqrt{T}})$ and $O(\frac{1}{T})$.
\par From \eqref{E68}, because of the $Err_T$ does not decrease with the increase of the sampling times $T$, normalized distance of KF-QSE cannot further reduce the normalized distance, resulting in the accuracy of the KF-QSE cannot continue to increase. But based on \eqref{E39}, $D(\rho_k,\rho_k^*)$ of the OPD-ADMM going to keep decreasing. When $k>100$, the $D(\rho_k,\rho_K^*)$ of KF-QSE is always in a state of oscillation, while the $D(\rho_k,\rho_k^*)$ of OPG-ADMM is still decreasing. The normalized distance of the two algorithms are close at  $k=480$, and the normalized distance of the two algorithms are 0.0206 and 0.0219.
When the sampling time comes to $k=1000$, the normalized distance of two algorithm are 0.0205 and 0.0091, it show that the reconstruction error of OPG-ADMM is much smaller than KF-QSE.
\par In summary, the KF-QSE can make the reconstruction density matrix $\rho_k$ converge to the real density matrix $\rho_k^*$ at a faster rate, but the KF-QSE always has an error term $Err_T$, resulting in the KF-QSE can not further improve the accuracy. For OPG algorithm, its convergence rate is slower than KF algorithm, but it can achieve higher precision by increasing the sampling time.
\section{Conclusion}
Based on the research of convergence rate theorems of online ADMM algorithm and online gradient descent algorithm, this paper analyzed the theoretical performance of two proposed online quantum state reconstruction algorithms: OPG-ADMM algorithm and KF-QSE algorithm. By proving the theorem, it has been found that KF algorithm has a faster convergence rate than OPG-ADMM algorithm, but at the same time, it has an error term, which can not reach a higher precision in the process of multiple iterations. Although the convergence rate of OPG-ADMM algorithm is low, it can achieve higher accuracy through multiple iterations.


\end{document}